\documentclass[11pt]{article}
\usepackage{moriond,epsfig}

\usepackage{amsmath,amssymb}

\def\be{\begin{equation}}
\def\ee{\end{equation}}
\def\bea{\begin{eqnarray}}
\def\eea{\end{eqnarray}}

\def\f{\frac}

\def\g{\gamma}

\newcommand{\xp}{x_{\mathbb P}}

\begin{document}
\vspace*{4cm}
\title{NUCLEAR DIFFRACTIVE STRUCTURE FUNCTIONS AT HIGH ENERGIES}

\author{C. MARQUET$^{1,2}$, H. KOWALSKI$^3$, T. LAPPI$^2$ and R. VENUGOPALAN$^4$}
\address{$^1$Department of Physics, Columbia University, New York, NY 10027, USA\\
$^2$Institut de Physique Th\'eorique, B\^at. 774, CEA/DSM/Saclay, 91191 Gif-sur-Yvette, France\\
$^3$Deutsches Elektronen-Synchrotron DESY, 22607 Hamburg, Germany\\
$^4$Physics Department, Brookhaven National Laboratory, Upton, NY 11973, USA}

\maketitle\abstracts{
A future high-energy electron-ion collider would explore the non-linear
weakly-coupled regime of QCD, and test the Color Glass Condensate (CGC) approach
to high-energy scattering. Hard diffraction in deep inelastic scattering off nuclei
will provide many fundamental measurements. In this work, the nuclear diffractive
structure function $F_{2,A}^D$ is predicted in the CGC framework, and the features
of nuclear enhancement and suppression are discussed.
}

\section{Introduction}

The understanding of hard diffraction in electron-proton (e-p) deep inelastic 
scattering (DIS) has been a great theoretical challenge since diffractive
processes were observed at HERA, and shown to represent more than 15\%
of all events. It was quickly understood that hard diffraction probes QCD in a
different way than hard inclusive measurements, for instance, unitarization should
be an important ingredient \cite{gbwdiff} of the description of diffractive cross-sections
at high energies, making those observables ideal places to look for manifestations of non-linear
saturation effects in QCD, such as geometric scaling.\cite{gsdiff}

In this work, hard diffraction in electron-nucleus (e-A) collisions is considered
within the IPsat model,\cite{kt} corresponding to the classical limit of the Color Glass Condensate approach.\cite{cgcrev} This effective theory of QCD at high partonic density is the most natural framework to describe the saturation phenomenon, and therefore to study e-A scattering at high energies, in particular diffractive observables. Here we shall focus on the nuclear diffractive structure function $F_{2,A}^D.$

Let us recall the kinematics of diffractive DIS: $\g^*A\!\rightarrow\!XA.$ With a momentum transfer $t\!\leq\!0,$ the proton/nucleus gets out of the $\g^*\!-\!A$ collision intact, and is separated by a rapidity gap from the other final-state particles whose invariant mass we denote $M_X.$ The photon virtuality is denoted $Q^2,$ and the $\g^*\!-\!A$ total energy $W.$ It is convenient to introduce the following variables: $x\!=\!Q^2/(Q^2\!+\!W^2),$
$\beta\!=\!Q^2/(Q^2\!+\!M_X^2)$ and $\xp\!=\!x/\beta.$ The size of the rapidity gap is
$\ln(1/\xp).$

The diffractive structure function is expressed as a function of $\beta,$ $\xp,$ $Q^2,$ and $t,$ and we will only consider the $t-$integrated structure function $F_2^{D,3}.$ While at large values of $\xp$ and $Q^2,$ the leading-twist collinear factorization is appropriate to describe hard diffraction off protons, this is not the case at small $\xp$ or off nuclei, as higher twists are enhanced by $\sim(A/\xp)^{0.3}.$ In this situation, the dipole picture is better suited to address the problem. It naturally incorporates the description of both inclusive and diffractive events into a common theoretical framework:\cite{dipdiff} the same dipole-nucleus scattering amplitudes, which can be computed treating the nucleus as a CGC, enter in the formulation of the inclusive and diffractive cross-sections.

\section{Diffractive structure functions in the dipole picture}

In our approach, $F_2^D\!=\!F_T^{q\bar q}\!+\!F_L^{q\bar q}\!+\!F_T^{q\bar qg}$ where the different pieces correspond to transversely (T) or longitudinally (L) polarized photons dissociating into a $q\bar q$ or $q\bar qg$ final state. For instance, the $q\bar q$
contributions are
\bea
\xp F_T^{q\bar q}(\beta,\xp,Q^2)\!\!&\!\!=\!\!&\!\!
\f{N_c Q^4}{8\pi^3\beta}\sum_f e_f^2\int_0^1 dz\ \Theta(\kappa_f^2) 
z(1\!-\!z)\left[f_T(z)\varepsilon_f^2(z)I_1(\kappa_f,\epsilon_f)
\!+\!m_f^2I_0(\kappa_f,\epsilon_f)\right],\\
\xp F_L^{q\bar q}(\beta,\xp,Q^2)\!\!&\!\!=\!\!&\!\!
\f{N_c Q^6}{8\pi^3\beta}\sum_f e_f^2\int_0^1 dz\ \Theta(\kappa_f^2)
z(1\!-\!z)f_L(z)I_0(\kappa_f,\epsilon_f)\ ,
\label{qq}\eea
with
\be
\varepsilon_f^2(z)\!=\!z(1\!-\!z)Q^2\!+\!m_f^2\ ,\hspace{0.2cm}
\kappa_f^2(z)\!=\!z(1\!-\!z)M_X^2\!-\!m_f^2\ ,\hspace{0.2cm}
f_T(z)\!=\!z^2\!+\!(1\!-\!z)^2\ ,\hspace{0.2cm}
f_L(z)\!=\!4z^2(1\!-\!z)^2\ .
\ee
The $\xp$ dependence comes in the functions $I_\lambda$ from $N_A(r,b,\xp),$ the $q\bar q$ dipole-nucleus scattering amplitude:
\be
I_\lambda(\kappa,\epsilon)\!=\!\int d^2b
\left[\int_0^\infty\!rdr J_\lambda(\kappa r)K_\lambda(\epsilon r)N_A(r,b,\xp)\right]^2
\label{ilambda}\ee
where $J_\lambda$ and $K_\lambda$ are Bessel functions. In formula (\ref{ilambda}), the integration variables $r$ and $b$ are the $q\bar q-$dipole transverse size and its impact parameter.

In principle, it is justified to neglect final states containing gluons, because these are suppressed by extra powers of $\alpha_s.$ However, for small values of $\beta$ or large values of $Q^2,$ the $q\bar q$ pair will emit soft or collinear gluons whose emissions are accomponied by large logarithms $\ln(1/\beta)$ or $\ln(Q^2)$ which compensate the factors of $\alpha_s.$ 
In those situations, multiple gluons emissions should be resummed; in practice, including the
$q\bar qg$ final state is enough to describe the HERA data. In both the small$-\beta$ and
large$-Q^2$ limits, this can be done within the dipole picture. An implementation of the
$q\bar qg$ contribution $F_T^{q\bar qg}$ that correctly reproduces both limits was recently proposed,\cite{me} while at large $\beta$ and small $Q^2,$ the $q\bar q$ contributions
(\ref{qq}) dominate. The formulae that we shall use can be found in this work.\cite{me}

\section{The dipole-nucleus scattering amplitude}

We shall use the IPsat parametrization to describe the dipole-nucleus scattering amplitude:
\be
N_A(r,b,x)=1-e^{-r^2F(r,x)\sum_{i=1}^A T_p(b-b_i)}\ ,\hspace{0.5cm}
F(x,r^2)=\frac{\pi^2}{2N_c}\alpha_s(\mu_0^2\!+\!C/r^2)xg(x,\mu_0^2+C/r^2)\ .
\label{ipsat}\ee
This is a model of a nucleus whose nucleons interact independently. Indeed, $N_A$ is obtained from $A$ dipole-nucleon amplitudes $N_p\!=\!1\!-\!\exp[-r^2F(r,x)T_p(b)]$ assuming that the probability
$1\!-\!N_A$ for the dipole not to interact with the nucleus is the product of the probabilities
$1\!-\!N_p$ for the dipole not to interact with the nucleons. This assumption is not consistent with the CGC quantum evolution, which sums up nonlinear interactions between the nucleons. However, the classical limit (\ref{ipsat}) of the dipole-CGC scattering amplitude can be thought of an initial condition. Note that in the small $r$ limit, one has $N_A=\sum_i N_p,$ and there is no leading twist shadowing.

In (\ref{ipsat}), $T_p(b)\!\propto\!\exp[-b^2/(2 B_G)]$ is the impact parameter profile function in the proton with $\int d^2b\ T_p(b)=1,$ and $F$ is proportional to the DGLAP evolved gluon distribution. The parameters $\mu_0,$ $C,$ and $B_{\rm G}$ (as well as two other parameters characterising the initial condition for the DGLAP evolution) are fit to
reproduce the HERA data on the inclusive proton structure function $F_2.$ The diffractive proton structure function $F_2^D$ is well reproduced \cite{us} after adjusting $\alpha_s=0.14$ in the
$q\bar qg$ component. Vector-meson production at HERA is also well described.\cite{kmw}

We introduced in (\ref{ipsat}) the coordinates of the individual nucleons $\{b_i\},$ they are distributed according to the Woods-Saxon distribution $T_A(b_i),$ which means that to compute
an observable, one has to perform the following average
\be
\langle\mathcal{O}\rangle_N \equiv \int \left(\prod_{i=1}^A d^2b_i 
T_A(b_i)\right)\mathcal{O}(\left\{b_i\right\})\ .
\label{wsavg}\ee
The Woods-Saxon parameters are measured from the electrical charge distribution and no additional parameters are introduced. The dipole cross-sections obtained in this manner give a good agreement \cite{klv} with the small$-x$ NMC data on the nuclear structure function
$F_{2,A}.$ We shall now use this parametrization of $N_A$ to predict the nuclear diffractive structure function $F_{2,A}^D.$

Note that performing the average (\ref{wsavg}) at the level of the amplitude, meaning calculating
$\langle N_A\rangle_N^2$ in (\ref{qq}), imposes that the nucleus is intact in the final state, it hasn't broken up. By contrast, when performing the average at the level of the cross-section, meaning calculating $\langle N_A^2\rangle_N$ in (\ref{qq}), one allows the nucleus to break up into individual nucleons, which will typically happen when the momentum transfer is bigger than the inverse nuclear radius. In what follows, we shall refer to those two possibilities as ``non breakup'' (also known as coherent diffraction) and ``breakup'' cases (coherent+incoherent diffraction).

\begin{figure}[t]
\begin{minipage}[t]{79mm}
\centerline{\epsfxsize=7.5cm\epsfbox{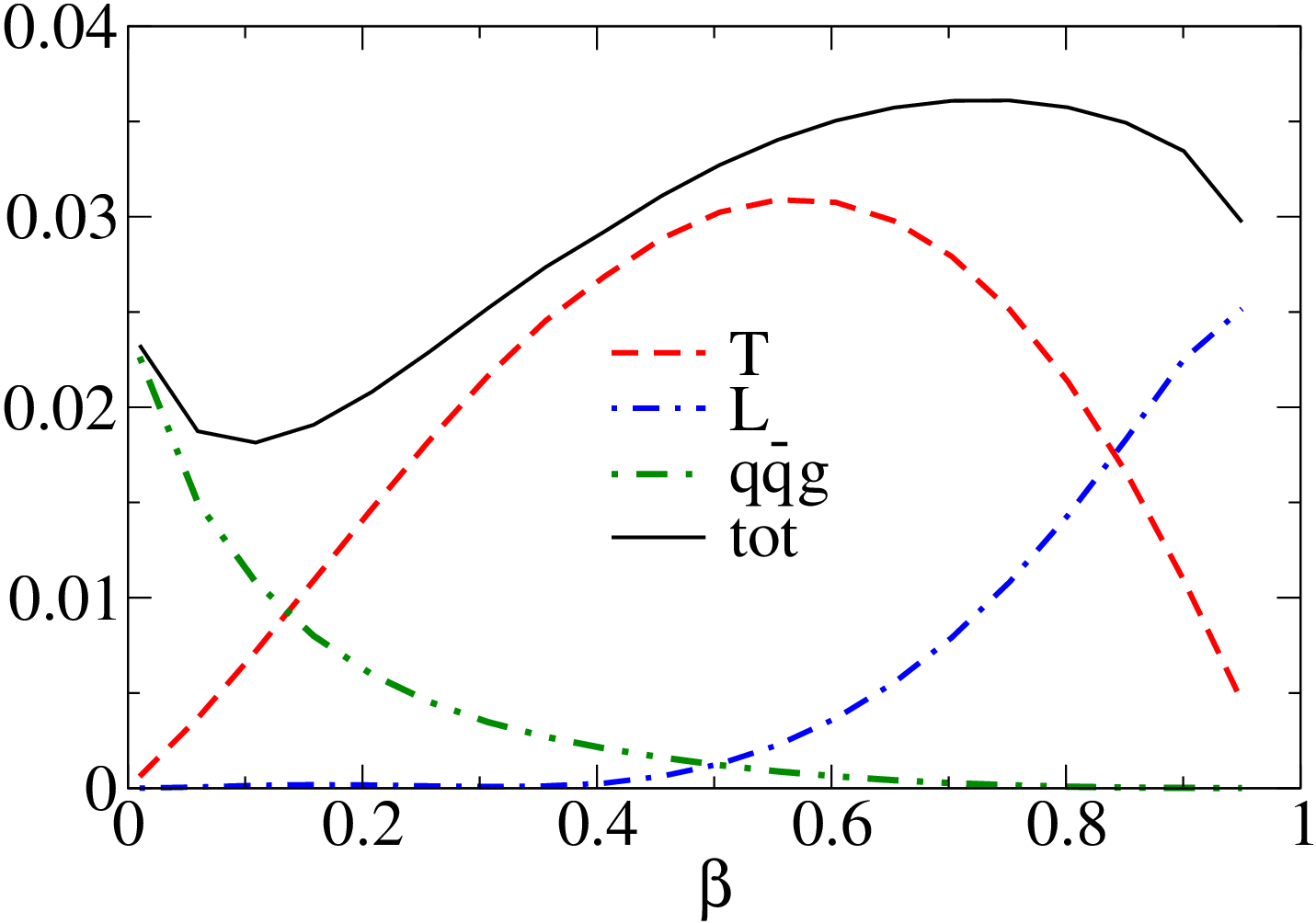}}
\end{minipage}
\hspace{\fill}
\begin{minipage}[t]{79mm}
\centerline{\epsfxsize=7.5cm\epsfbox{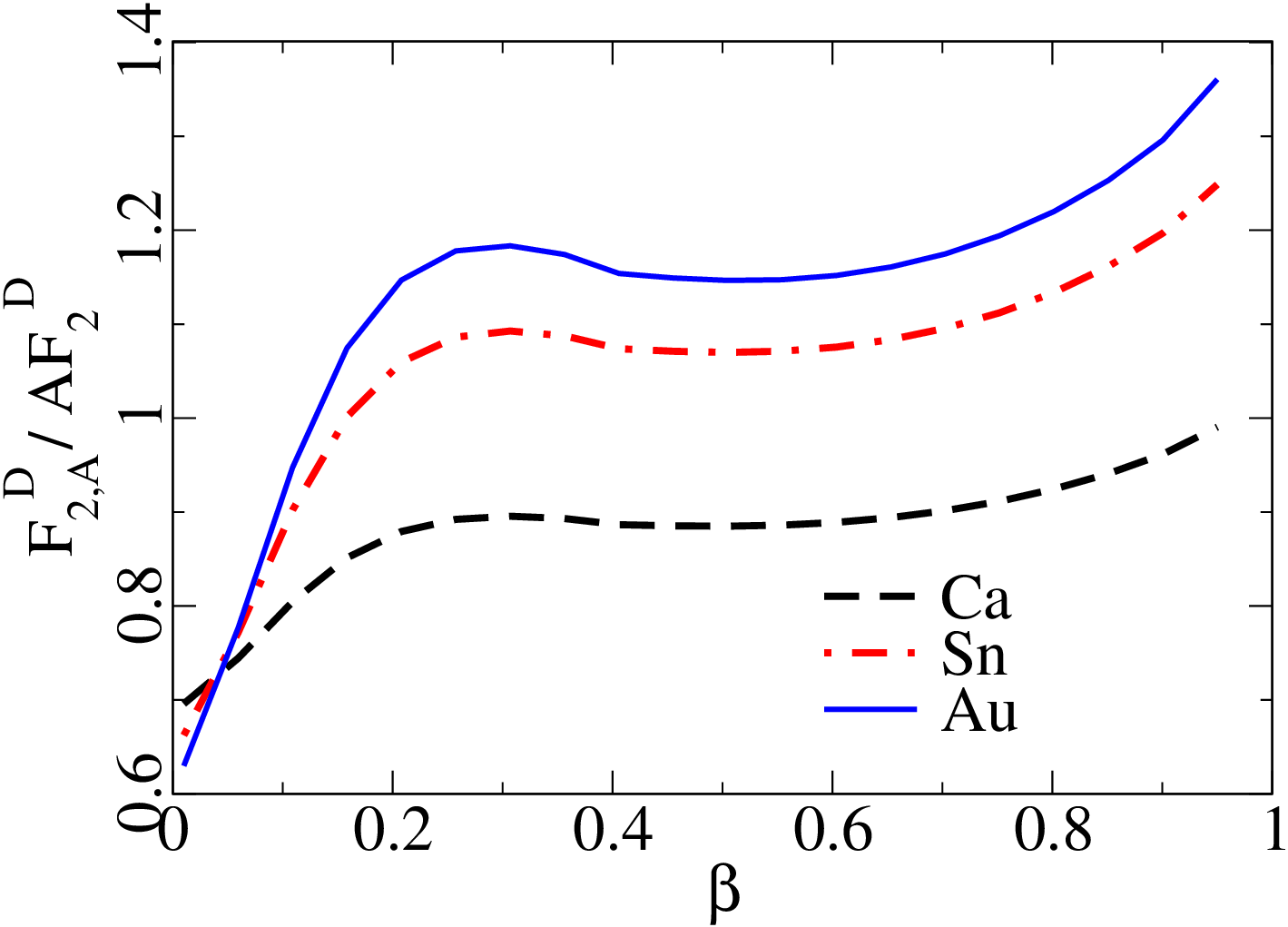}}
\end{minipage}
\caption{Left plot: $\beta$-dependence of the different contributions to the proton
diffractive structure function $F_{2,p}^D.$ Right plot: the ratio $F_{2,A}^D/(AF_{2,p}^D)$ as a function of $\beta$ for Ca, Sn and Au nuclei. In both cases, results are for the ``non breakup'' case, and at $Q^2=5\ \mbox{GeV}^2$ and $\xp=0.001$.}
\end{figure}

\section{Nuclear enhancement and suppression of $F_2^D$}

In Figure 1, the $\beta$ dependence of the diffractive structure function is displayed for
$Q^2=5\ \mbox{GeV}^2$ and $\xp=0.001.$
On the left plot, the hierarchy of the different contributions is analysed in the case of
$F_{2,p}^D.$ The dominant contribution is: the $q\bar q g$ component for
values of $\beta\!<\!0.1,$ the longitudinally polarized $q\bar q$ component for values of
$\beta\!>\!0.9,$ and the transversely polarized $q\bar q$ component for intermediate values.
In the case of $F_{2,A}^D,$ this separation is still true but the $q\bar q$ and $q\bar q g$ components behave differently as a function of $A.$ The $q\bar{q}$ components are enhanced compared to $A$ times the proton diffractive structure functions while the $q\bar{q}g$ component, on the contrary, is suppressed for nuclei compared to the proton (the $Q^2$ and $\xp$ dependence
of these effects will be discussed shortly).

This leads to a nuclear suppression of the diffractive structure function in the 
small $\beta$ region, and to an enhancement at large $\beta.$ This is illustrated
by the right plot of Figure 1, where the ratio $F_{2,A}^D/(AF_{2,p}^D)$ is shown as a function
of $\beta$ for different nuclei (for the ``non breakup'' case). The net result of the
different contributions is that $F_{2,A}^D/A$, for a large $\beta$ range down to 0.1, is close to
$F_{2,p}^D,$ and is increasing with $A.$

In Figure 2, for the Au nucleus case, the ratios $F_{2,A}^D/(AF_{2,p}^D)$ of individual contributions are analysed (for values of $\beta$ at which they are dominant).
Comparisons between the ``breakup'' and ``non breakup'' cases are made, as functions of $Q^2$ (left plot) and $\xp$ (right plot). For the $q\bar q g$ component, the nuclear suppression is almost constant (the suppression slightly decreases with $Q^2$). For the $q\bar q$ components, the enhancement becomes bigger with increasing $Q^2$ and $\xp.$ The result for the total diffractive cross-section in e-A scattering is that it decreases more slowly with increasing $Q^2$ or $\xp$ compared to the e-p case. Finally, cross sections in the ``non breakup'' case are about 15\%
lower than in the ``breakup'' case.

Comparing with other approaches, we obtain similar features. We notice one interesting difference with the results obtained using diffractive parton distributions modified by leading twist shadowing:\cite{fgs} even at large $\beta,$ it is found that $F_{2,A}^D/A$ is suppressed compared to $F^D_{2,p}$ as a function of $Q^2.$ This could be tested with measurements at a future electron-ion collider \cite{eic} where diffraction will be an important part of a rich program. A typical nuclear enhancement of diffraction, for a Au nucleus, is a factor of $\sim\!1.2.$ Combining this with the typical nuclear suppression in the inclusive case ($\sim\!0.8,$ see
\cite{klv}), we expect the fraction of diffractive events to be increased by a factor of
$\sim\!1.5$ compared to the proton, meaning 25\%
at an e-A collider.

\begin{figure}[t]
\begin{minipage}[t]{79mm}
\centerline{\epsfxsize=7.4cm\epsfbox{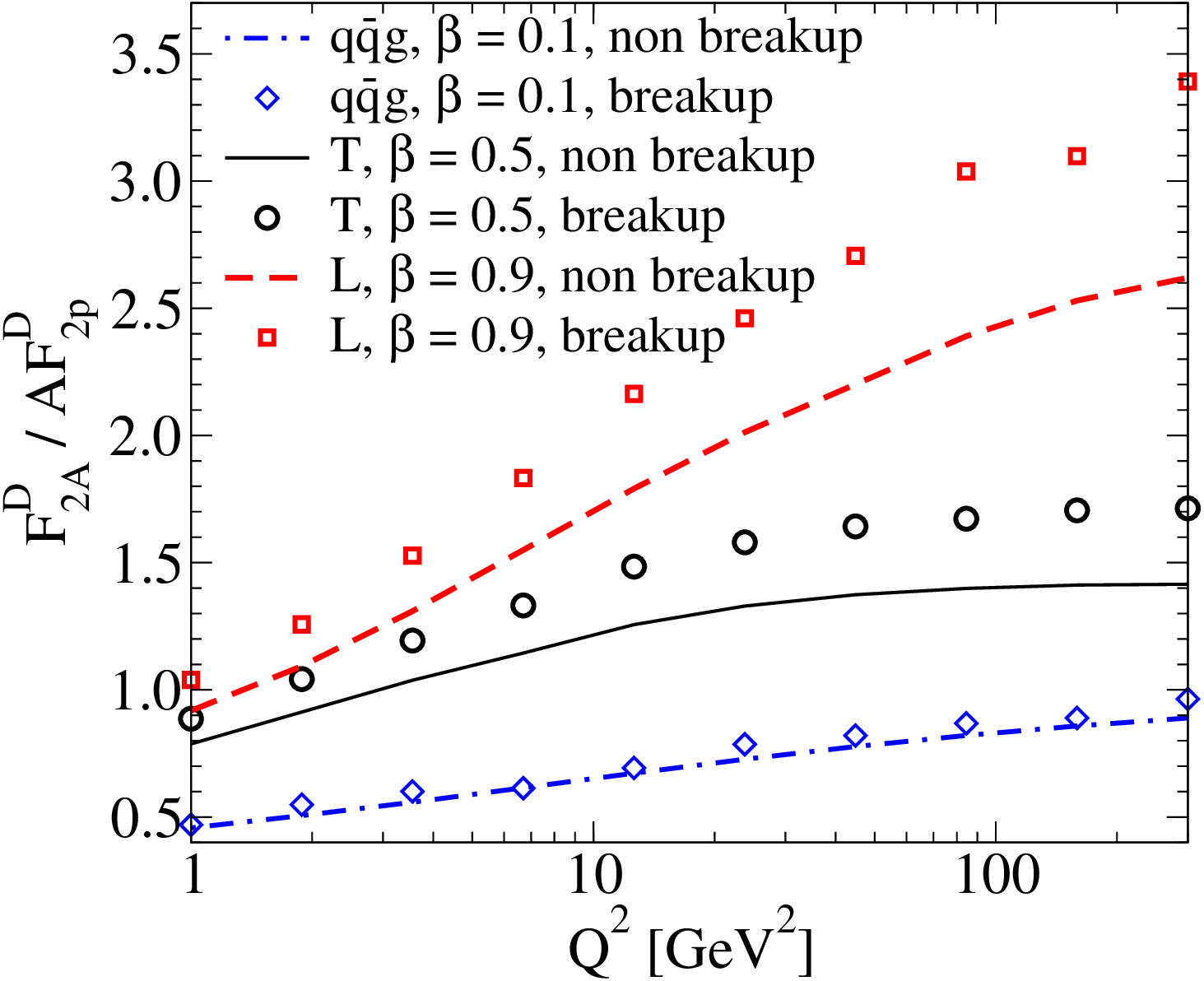}}
\end{minipage}
\hspace{\fill}
\begin{minipage}[t]{79mm}
\centerline{\epsfxsize=7.6cm\epsfbox{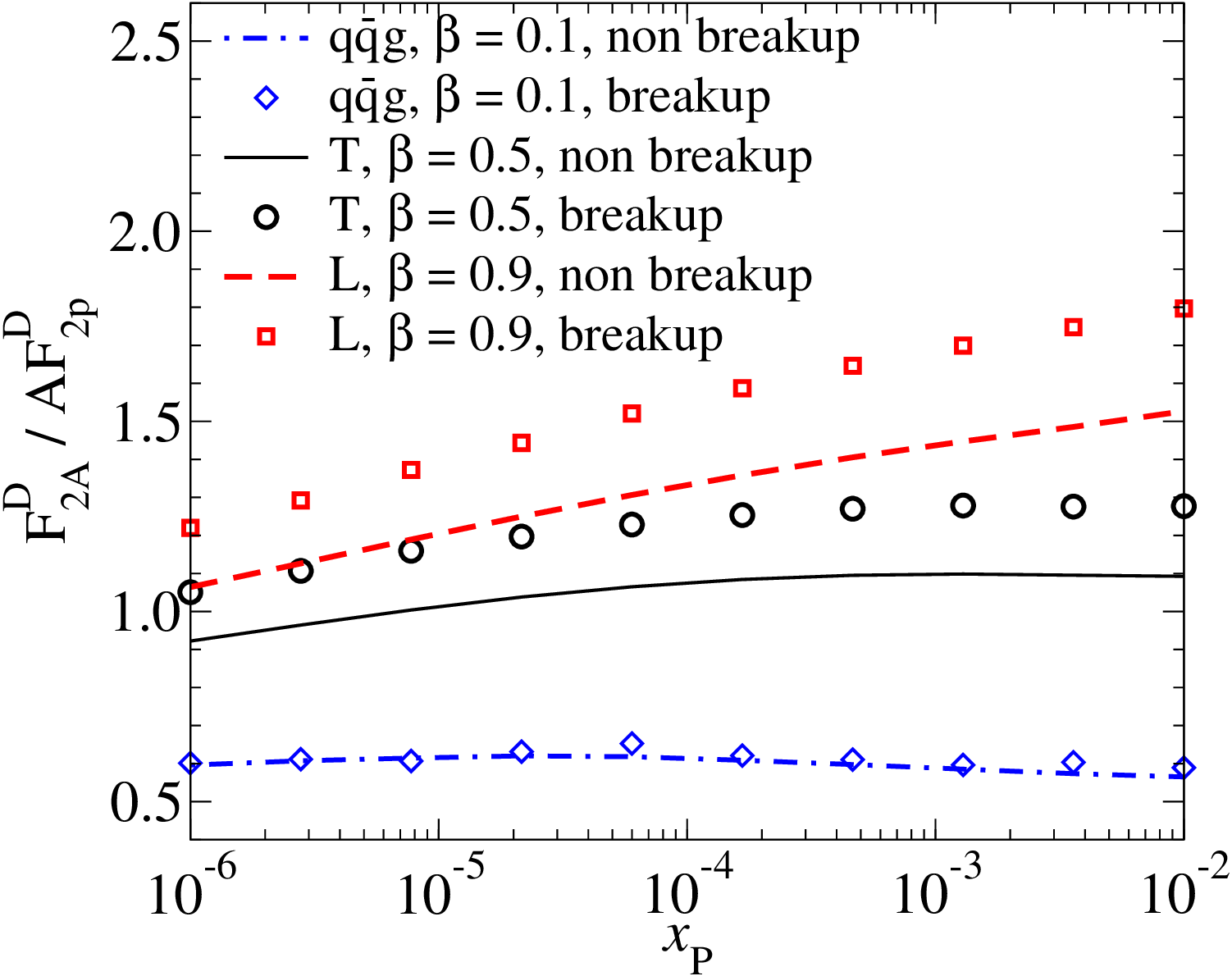}}
\end{minipage}
\caption{The ratios $F_{2,A}^{D,x}/(AF_{2,p}^{D,x})$ of the different components
($x\!=\!q\bar qg,\ q\bar qT,\ q\bar qL$) of the diffractive structure function for both ``breakup''
and ``non breakup'' cases. Left plot: as a function of $Q^2$ for $\xp\!=\!0.001.$ 
Right plot: as a function of $\xp$ for $Q^2\!=\!5\ \mbox{GeV}^2.$
In both cases, results are for Au nuclei and the different components
are evaluated where they are dominant: at $\beta\!=\!0.1$ for $q\bar{q}g$,
$\beta\!=\!0.5$ for $q\bar qT$ and $\beta\!=\!0.9$ for $q\bar qL.$}
\end{figure}

\end{document}